\documentclass[aps,pra,showkeys,twocolumn,superscriptaddress]{revtex4-2}
\usepackage{array}
\usepackage{booktabs}
\usepackage{tabu}
\usepackage{dcolumn}
\usepackage{amsmath}
\usepackage{amsfonts}
\usepackage{float}
\usepackage{amssymb}
\usepackage{graphicx,color}
\usepackage[colorlinks={true}]{hyperref}
\hypersetup{colorlinks=true,linkcolor=red,citecolor=blue,urlcolor=blue}
\usepackage{graphicx}
\usepackage{subfigure}
\usepackage{graphicx}
\usepackage{dcolumn}
\usepackage{bm}
\usepackage{pstricks}
\usepackage{braket}
\usepackage{orcidlink}

\def\be{\begin{equation}}
  \def\ee{\end{equation}}
\def\bea{\begin{eqnarray}}
\def\eea{\end{eqnarray}}
\def\f{\frac}
\def\n{\nonumber}
\def\l{\label}
\def\p{\phi}
\def\o{\over}
\def\R{\rho}
\def\pa{\partial}
\def\om{\omega}
\def\na{\nabla}
\def\P{\Phi}
\begin{document}

\title{Quantum speed limit and non-Markovianity in structured environments} 

\author{Maryam Hadipour \orcidlink{0000-0002-6573-9960}}
\affiliation{Faculty of Physics, Urmia University of Technology, Urmia, Iran}

\author{Soroush Haseli \orcidlink{0000-0003-1031-4815}}\email{soroush.haseli@uut.ac.ir}
\affiliation{Faculty of Physics, Urmia University of Technology, Urmia, Iran}
\affiliation{School of Physics, Institute for Research in Fundamental Sciences (IPM),  P.O. Box 19395-5531, Tehran, Iran}

\author{Saeed Haddadi \orcidlink{0000-0002-1596-0763}} \email{haddadi@semnan.ac.ir}
\affiliation{Faculty of Physics, Semnan University, P.O. Box 35195-363, Semnan, Iran}

\date{\today}
\def\be{\begin{equation}}
  \def\ee{\end{equation}}
\def\bea{\begin{eqnarray}}
\def\eea{\end{eqnarray}}
\def\f{\frac}
\def\n{\nonumber}
\def\l{\label}
\def\p{\phi}
\def\o{\over}
\def\R{\rho}
\def\pa{\partial}
\def\om{\omega}
\def\na{\nabla}
\def\P{$\Phi$}

\begin{abstract}
We investigate the relationship between quantum speed limit time and the non-Markovianity of an atom in structured environments.  We show that there exists an inverse relation between them, which means that the non-Markovian feature of the quantum process leads to speedup of the process. Our results might shed light on the relationship between the speedup of quantum evolution and the backflow of information from the environment to the system. 
\end{abstract}

\maketitle

\section{Introduction}
How to accelerate the evolution of quantum systems is one of the most important questions in quantum mechanics theory. Based on this theory, quantum speed limit (QSL) quantifies the minimum evolution time for a quantum system to pass through a predetermined distance  \cite{Deffner2017,Hilgevoord2002,Deffner2017p,Pfeifer1993,Bukov2019,Funo2019,Hegerfeldt2013,Garcia-Pintos2019}.  It refers to the shortest possible time required for a quantum system to evolve from an arbitrary initial state to a target final state \cite{Garcia-Pintos2019p,Kobe1994,Jones2010,Xu2016,Deffner2013a,Shao2020,Russell2014, Hu2020}.  QSL time and its applications have been extensively studied experimentally and theoretically \cite{Cheneau2012,Campo2021,Taddei2013,Escher2011,Lam2021}. Moreover, QSL time plays a particularly important role in the development of quantum information protection and quantum optimal theory as well \cite{Poggi2019,Caneva2009,Zhang2021,Marvian2015}. In recent years, there has been a particular focus on utilizing the advantages of QSL time in quantum batteries  \cite{Bai2020,Campaioli2017,Hovhannisyan2013,Binder2015}.
  A deeper knowledge of QSL has led to the expansion of the speed limit to classical systems \cite{OkuyamaMand2018,Shanahan2018,Bolonek-Lason2021}. There is a strong correlation between the vanishing QSL times and the emerging classical behavior due to the reduced uncertainty in quantum observables \cite{Poggi2021}. In Ref. \cite{arXiv:2004.03078}, the application of QSL time to quantum resource theories has been investigated. In the beginning, only isolated systems were studied in accordance with the concept of QSL time \cite{Mandelstam1991,Margolus1998,Levitin2009}. 
  
  Initially, Mandelstam and Tamm (MT) introduced the minimal time for evolution from an initial to an orthogonal state for a closed quantum system as $\tau \geqslant \frac{\pi \hbar}{2 \Delta E}$, where $(\Delta E)^2=\left\langle\psi\left|H^2\right| \psi\right\rangle-(\langle\psi|H| \psi\rangle)^2$ is the
energy variance of the system \cite{Mandelstam1991}. This bound is known as MT bound. Another bound related to the mean energy $E=\langle \psi \vert H \vert \psi \rangle$ was then derived by Margolus and Levitin (ML) as $\tau \geq \frac{\pi \hbar}{2 E}$, where it is assumed that the ground state energy $E_0$ is equal to zero \cite{Margolus1998}. This bound is known as ML bound. It is possible to implement a unified QSL bound for the unitary evolution of closed quantum systems by combining both ML and MT bounds as \cite{Levitin2009}
\begin{equation}
\tau \geq \max \left\{ \frac{\pi \hbar}{2 \Delta E}, \frac{\pi \hbar }{2 E}\right\}. 
\end{equation} 

The ML and MT bounds have been expanded and improved using a wide range of metrics, including trace distance, Bures angle, relative purity, and many other metrics \cite{Jozsa1994,Uhlmann1976,Luo2004,Wu2020}.  The distance among generalized Bloch vectors for both unitary and non-unitary processes can also be used to obtain tight QSL bounds \cite{Deffner2013,Wu2018,Cai2017,del2013,Ektesabi2017,Liu2016}.

It should be noted that in the real world, quantum systems are open and their evolution is non-unitary \cite{Breuer2007}. A variety of approaches can be used to analyze the behavior of open systems, including the master equation formalism and the quantum trajectory method. The master equation is a central tool for studying the dynamics of open systems. It provides a mathematical description of how the system's density matrix changes over time. The nature of an interaction between a quantum system and its environment can influence its dynamics and evolution.  The system-environment interaction is most often analyzed from both the non-Markovian and Markovian perspectives \cite{Breuer2007}. In a Markovian regime, the evolution of the system is purely determined by its current state. In this case, the system-environment interactions occur over a much shorter timescale than the significant system evolution.  However, in a non-Markovian regime, system-environment interactions are longer in time than the intrinsic evolution of the system \cite{Davies1976,Wolf2008,Breuer2009,Rivas2010,Luo2012,Zeng2011,He2017,Zhang2014}. An important feature of non-Markovian regimes is that the interaction behavior of the system with its environment in the past determines the evolution of the system in the future.
Due to the unavoidable interaction between a quantum system and its environment, the decoherence effect caused by this coupling can have a significant effect on quantum statistics.

The ratio between the QSL time and the actual time quantifies the possibility of speeding up the quantum evolution. QSL bounds are saturated with no potential for speedup when the ratio equals one. However, the QSL bound is unsaturated if the ratio is less than one, and there exists potential for speeding up the evolution of the dynamical system. In Ref. \cite{Deffner2013}, it has been shown that the non-Markovian effect can accelerate quantum evolution \cite{Deffner2013}. Both theoretically and experimentally, non-Markovianity regulation has been proposed as a mechanism for regulating dynamical speedup. Moreover, it has been shown that non-Markovian effects induce unsaturated QSL bounds in open dynamics \cite{Cimmarusti2015,Zhang2015,Cianciaruso2017,Sun2015,Xu2019,Xu2014}. It is important to note that dynamical speedup in this way depends very much on the feedback from the environment. Therefore, it will be essential to investigate other sources that can generate unsaturated QSL bounds besides non-Markovianity as a source of unsaturated QSL bounds \cite{Teittinen2019,Berrada2018,Shahri2023,Gholizadeh2023,Hadipour2023}.

Motivated by this, we investigate the possibility of quantum speedup using a convenient cavity-based engineered environment. We show how the Markovian environments consisting of two cavities can accelerate the dynamics of an artificial atom interacting with a pseudomode. In the considered model, the coupling between the pseudomodes, atom-pseudomode coupling, and detuning between atom and pseudomode can have significant impacts on QSL time and non-Markovianity.  

The work is organized as follows. First, we present a physical model and its solution in Sec. \ref{model}.  Then, we discuss the QSL time and non-Markovianity for the considered model in Sec. \ref{QSLs}. Next, we turn to the results and discussion in Sec. \ref{Results}. Finally, we provide a brief conclusion of the most important results in Sec. \ref{conclusion}.

 \section{Model and pseudomode method}\label{model}
\begin{figure}[t]
    \centering
  \includegraphics[width = 1\linewidth]{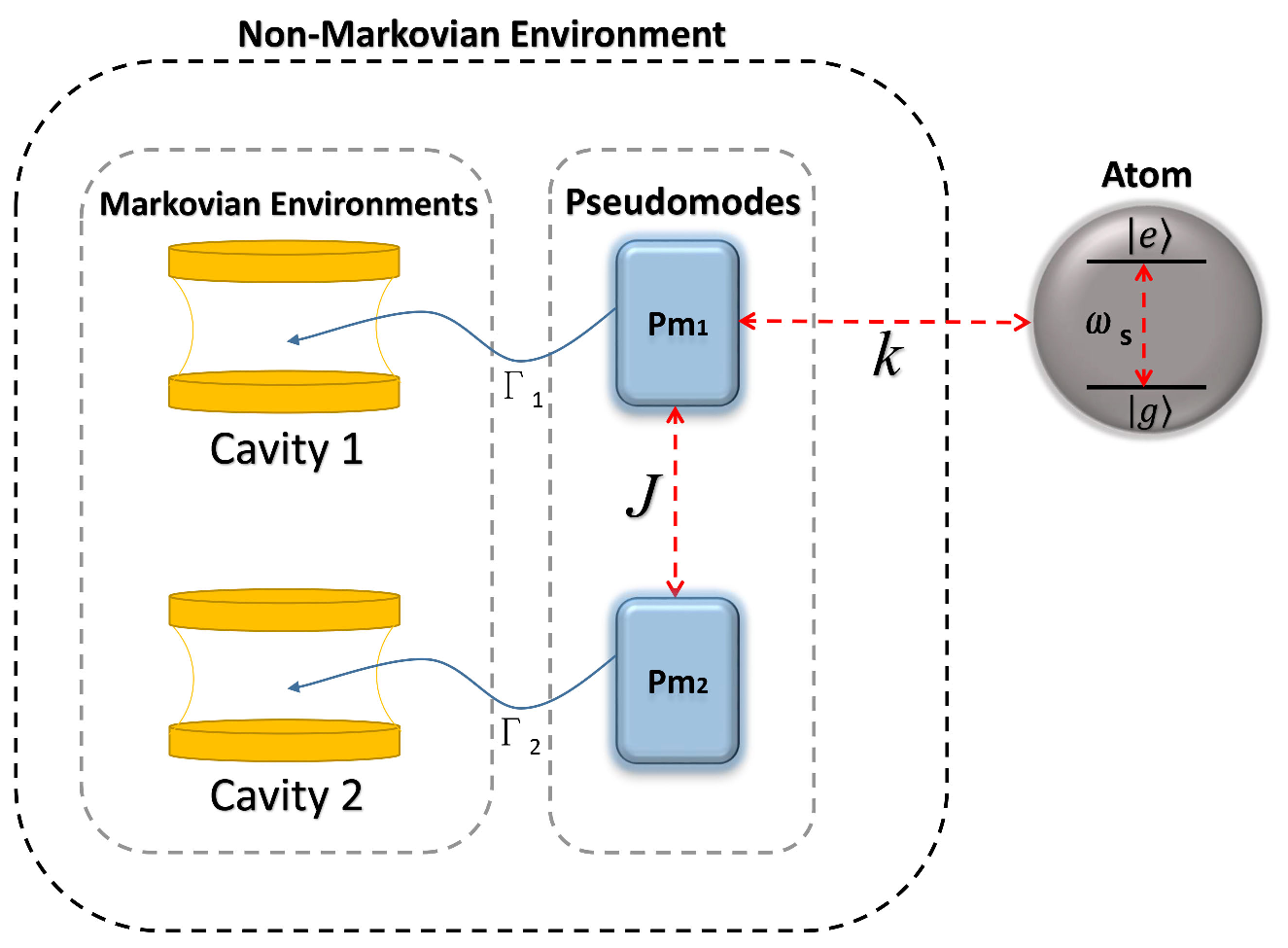}
    \centering
    \caption{Schematic representation of the model. Two-level atom interacts with pseudomode $1$ (coupling strength $k$) which leaks into cavity $1$ via dissipation rate $\Gamma_1$, and pseudomode $1$ is coupled to pseudomode $2$ with coupling strength $J$ which the latter leaks into cavity $2$ via dissipation rate $\Gamma_2$.
}\label{Fig1}
\end{figure}

The considered model consists of a two-level atom interacting with a Lorentzian structured environment. Figure \ref{Fig1} shows the schematic representation of the scheme. For this model, the Hamiltonian can be expressed as follows \cite{Mazzola2009,Xu2021}
\begin{equation}\label{Hamiltonian}
\begin{aligned}
\hat{H}= & \frac{\omega_s}{2}  \hat{\sigma}_z+\omega_1 \hat{a}_1^{\dagger} \hat{a}_1+\omega_2 \hat{a}_2^{\dagger} \hat{a}_2\\
&+k\left(\hat{a}_1^{\dagger} \hat{\sigma}_{-}+\hat{a}_1 \hat{\sigma}_{+}\right) 
 +J\left(\hat{a}_1^{\dagger} \hat{a}_2+\hat{a}_1 \hat{a}_2^{\dagger}\right),
\end{aligned}
\end{equation}
where $\hat{\sigma}_z=|e\rangle\langle e|-| g\rangle\langle g|$ is the ordinary Pauli operator in $z$-direction of atom, $\omega_s$ is the transition frequency of the atom, $\sigma_{+}$($\sigma_-$) is raising (lowering) operator of the atom,
and $\omega_1$ and $\omega_2$ are the frequencies of the modes inside the Markovian environments shown by cavity $1$ and cavity $2$, respectively. Let us take $\omega_1=\omega_2=\omega$ and set $\omega_0=\omega + \delta$, where $\delta$ is atom-pseudomode detuning.  Besides $\hat{a}_1$($\hat{a}^{\dag}_1$) and $\hat{a}_2$($\hat{a}^{\dag}_2$) are the annihilation (creation) operators. 

In what follows, we study the dynamics using the pseudomode theory \cite{pseudomode1,pseudomode2,pseudomode3,pseudomode4}. This approach relies on the relationship between the atom dynamics and the shape of the spectral distribution of the environment. The significance of precise approaches is of current interest due to the experimental realization of quantum systems. As shown in Fig. \ref{Fig1}, the environment can be depicted by two nondegenerate pseudomodes that leak into the Markovian environments (cavities $1$ and $2$) with dissipation rates $\Gamma_1$ and $\Gamma_2$. Here, the two-level atom only interacts with the first pseudomode Pm$_1$ (the strength of the coupling $k$) which is in turn coupled to the second pseudomode Pm$_2$ (the strength of the coupling $J$).   

The dynamics of the whole system for a Lorentzian spectral distribution can be obtained by the following exact pseudomode master equation 
\begin{equation}\label{masterequation}
\begin{aligned}
\dot{\rho}(t)= & -i[\hat{H}, \rho(t)] \\
& -\sum_{n=1, 2} \frac{\Gamma_n}{2}\left[a_n^{\dagger} a_n \rho(t)-2 a_n \rho(t) a_n^{\dagger}+\rho(t) a_n^{\dagger} a_n\right].
\end{aligned}
\end{equation}

The above master equation represents the coherent interaction between the two-level atom and two pseudomodes in the presence of the pseudomodes decay caused by the interaction with Markovian environments.

Here, it is assumed that the atom is initially in an excited state $\vert e_a \rangle$ while both pseudomodes Pm$_\textmd{1}$ and Pm$_\textmd{2}$  are in the ground state $\vert g_{\textmd{Pm}_1} g_{\textmd{Pm}_2} \rangle$, that is, the initial state of the overall system is assumed to be $\rho(0)=\vert e_a g_{\textmd{Pm}_1} g_{\textmd{Pm}_2} \rangle \langle e_a g_{\textmd{Pm}_1} g_{\textmd{Pm}_2} \vert $. Considering that there is only one excitation in the whole system including a two-level atom and two cavities, the state of the system at time $t$ can be written as
\begin{align}\label{solution}
\rho(t)=&(1-\eta(t))|\psi(t)\rangle\langle\psi(t)|\nonumber\\
&+\eta(t)| g_a g_{\textmd{Pm}_1} g_{\textmd{Pm}_2}\rangle\langle g_a g_{\textmd{Pm}_1} g_{\textmd{Pm}_2}|,
\end{align}
where $0 \leq \eta(t) \leq 1$ with $\eta(0)=0$, and we have 
\begin{align}
|\psi(t)\rangle=&c_1(t)|g_a e_{\textmd{Pm}_1} g_{\textmd{Pm}_2}\rangle+c_2(t)|g_a g_{\textmd{Pm}_1} e_{\textmd{Pm}_2}\rangle\nonumber\\
&+h(t)|e_a g_{\textmd{Pm}_1} g_{\textmd{Pm}_2}\rangle,
\end{align}
with $h(0)=1$ and $c_1(0)=c_2(0)=0$. For convenience, it is appropriate to consider the following unnormalized state as 
\begin{equation}
\begin{aligned}
|\widetilde{\psi}(t)\rangle  \equiv& \sqrt{1-\eta(t)}|\psi(t)\rangle\\
=&\tilde{c}_1(t)|g_a e_{\textmd{Pm}_1} g_{\textmd{Pm}_2}\rangle+\tilde{c}_2(t)|g_a g_{\textmd{Pm}_1} e_{\textmd{Pm}_2}\rangle\\
&+\tilde{h}(t)|e_a g_{\textmd{Pm}_1} g_{\textmd{Pm}_2}\rangle,
\end{aligned}
\end{equation}
where $\widetilde{h}(t)=\sqrt{1-\eta(t)} h(t)$ is the probability amplitude that the two-level atom is in an excited state while $\tilde{c}_n(t) \equiv \sqrt{1-\eta(t)}c_n(t)$ (with $n=1,2$) is the probability amplitude that the pseudomodes being in their excited states. Using the above unnormalized state, the density matrix of the total system can be rewritten as 
\begin{equation}
\rho(t)=|\widetilde{\psi}(t)\rangle\langle\widetilde{\psi}(t)|+\eta(t)|g_a g_{\textmd{Pm}_1} g_{\textmd{Pm}_2}\rangle\langle g_a g_{\textmd{Pm}_1} g_{\textmd{Pm}_2}|.
\end{equation}

The time-dependent amplitudes of $\tilde{h}(t)$, $\tilde{c}_1(t)$, and $\tilde{c}_2(t)$ can be obtained by solving the following coupled differential equations
\begin{equation}
\begin{aligned}
i \frac{d \widetilde{h}(t)}{d t} & =(\omega+\delta) \widetilde{h}(t)+k \tilde{c}_1(t), \\
i \frac{d \tilde{c}_1(t)}{d t} & =\left(\omega-\frac{i}{2} \Gamma_1\right) \tilde{c}_1(t)+k \widetilde{h}(t)+J \tilde{c}_2(t), \\
i \frac{d \tilde{c}_2(t)}{d t} & =\left(\omega-\frac{i}{2} \Gamma_2\right) \tilde{c}_2(t)+J \tilde{c}_1(t) .
\end{aligned}
\end{equation}

In the above set of coupled differential equations, the coefficients $\widetilde{h}(t)$, $\tilde{c}_1(t)$, and $\tilde{c}_2(t)$ can be obtained using the standard Laplace transformations. By knowing these,  the time-evolved density matrix of the two-level atom can be obtained by taking the partial trace over two pseudomodes, which we will provide in the next section.

\section{QSL time and non-Markovianity}\label{QSLs}

\subsection{QSL time} 
We first deal with QSL time using the method introduced in Ref. \cite{Wu2018} for open quantum systems.  A comprehensive bound of the QSL time for any arbitrary initial state can be defined as
 \begin{equation}
 \tau_{\textmd{QSL}}=\max \left\{\frac{1}{\Lambda_\tau^{o p}}, \frac{1}{\Lambda_\tau^{t r}}, \frac{1}{\Lambda_\tau^{h s}}\right\} \sin ^2[\Theta(\rho_0, \rho_\tau)] \operatorname{tr}\left[\rho_0^2\right],
 \end{equation}
where $\Theta(\rho_0, \rho_t)=\arccos \left(\sqrt{\frac{\operatorname{tr}[\rho_0 \rho_t]}{\operatorname{tr}\left[\rho_0^2\right]}}\right)$ is a function of relative purity between initial state $\rho_0$ and the state at arbitrary time $\rho_t$ that determined by the time-dependent non-unitary master equation $\dot{\rho}_t=\mathcal{L}_t (\rho_t)$ and 
\begin{equation}
\begin{aligned}
&\Lambda_\tau^{o p}=\frac{1}{\tau} \int_0^\tau d t\|\mathcal{L}_t (\rho_t)\|_{o p}, \\ &\Lambda_\tau^{tr}=\frac{1}{\tau} \int_0^\tau d t\|\mathcal{L}_t (\rho_t)\|_{tr}, \\ &\Lambda_\tau^{hs}=\frac{1}{\tau} \int_0^\tau d t\|\mathcal{L}_t (\rho_t)\|_{hs},
\end{aligned}
\end{equation} 
where $\|\mathcal{L}_t (\rho_t)\|_{o p}=\lambda_1$, $\|\mathcal{L}_t (\rho_t)\|_{tr}=\sum_i \lambda_i$ and $\|\mathcal{L}_t (\rho_t)\|_{hs}=\sqrt{\sum_i \lambda_i^2}$ are operator norm, trace norm and Hilbert-Schmidth norm, respectively. Also, $\lambda_i$'s are singular values of $\mathcal{L}_t (\rho_t)$ and $\lambda_1$ is the largest singular value of $\mathcal{L}_t (\rho_t)$.  When the denominator of the fraction is $\Lambda_\tau^{o p}$ and $\Lambda_\tau^{tr}$, we have a generalized ML type of QSL bound for open quantum systems, whereas when it is $\Lambda_\tau^{hs}$, we have MT type bound on the QSL time.  According to Ref. \cite{Wu2018}, the ML type bound based on the operator norm provides the tightest QSL bound.

Using the recently achieved bound for QSL time and a given actual driving time $\tau$, one can estimate the intrinsic speed of a dynamical evolution.  In the case of $\tau_{\textmd{QSL}}=\tau$, the further speedup is not possible and the speedup is not observed. In contrast, when $\tau_{\textmd{QSL}}<\tau$, this indicates the possibility of quantum dynamics speeding up. Based on the assumption that the two-level atom is initially in the excited state as mentioned in the previous section, the time-evolved density matrix of the atom takes the following form 
\begin{equation}\label{rohatom}
\rho_t= \vert \widetilde{h}(t) \vert^{2} \vert e \rangle\langle e \vert + (1-\vert \widetilde{h}(t) \vert^{2})\vert g \rangle \langle g \vert.
\end{equation}

So, QSL time based on the tightest bound is obtained as 
\begin{equation}\label{QSL}
\tau_{\textmd{QSL}}=\frac{1 -\vert \widetilde{h}(\tau) \vert^{2}}{\frac{1}{\tau}\int_0^{\tau} \vert \partial_t \vert \widetilde{h}(t) \vert^{2} \vert dt}.
\end{equation}

\subsection{Non-Markovianity}
During evolution, the backflow of information from the environment to the system is responsible for the non-Markovian feature of the process. For quantification, the degree of non-Markovianity can be defined as \cite{Breuer2009}
\begin{equation}\label{nong}
\mathcal{N}=\max _{\rho_{1}(0),~\rho_{2}(0)} \int_{\sigma>0} \sigma\left[t,~\rho_{1}(0),~\rho_{2}(0)\right] d t,
\end{equation}
in which
\begin{equation}
\sigma\left[t,~\rho_{1}(0),~\rho_{2}(0)\right]=\frac{d}{d t} D\left[\rho_1(t),~\rho_2(t)\right],
\end{equation}
where $D\left[\rho_1(t),~\rho_2(t)\right]=\frac{1}{2} \operatorname{Tr}\left|\rho_1(t)-\rho_2(t)\right|$ is the trace distance of a pair of states with the trace norm of an operator as $\vert \hat{A} \vert = \sqrt{\hat{A}\hat{A}^\dag}$. 

Note that when $\sigma\left[t,~\rho_{1}(0),~\rho_{2}(0)\right]>0$, we have the backflow of information from the environment to the system and so, the evolution is non-Markovian $\mathcal{N}\neq0$. However, if $\sigma\left[t,~\rho_{1}(0),~\rho_{2}(0)\right]<0$ then $\mathcal{N}=0$, meaning that the evolution is Markovian, information irreversibly flows from system to environment, and backflow does not occur. As a result, non-Markovianity indicates the total backflow of information from the environment and the system. 

In ref. \cite{Breuer2009}, it has been proved that the optimal pair of initial states that maximize the degree of non-Markovianity $\mathcal{N}$ are $\rho_1(0)=\vert e \rangle \langle e \vert$ and $\rho_2(0)=\vert g \rangle \langle g \vert$. Using the findings of the previous section and taking into account the mentioned optimal initial states, one can find $\mathcal{D}\left[\rho_1(t), \rho_2(t)\right]=|\widetilde{h}(t)|^2$. Therefore, the degree of non-Markovianity is obtained as 
\begin{equation}\label{non}
\mathcal{N}=\frac{1}{2}\left[\int_0^\tau \partial_t\vert \widetilde{h}(t)\vert^2  d t +\vert\widetilde{h}(\tau)\vert^2-1\right].
\end{equation}

\section{Results and discussion}\label{Results}
Now everything is ready to study the QSL time and non-Markovianity based on the analytical expressions \eqref{QSL} and \eqref{non} obtained in the previous section.  

\begin{figure}[t]
    \centering
  \includegraphics[width = 0.85\linewidth]{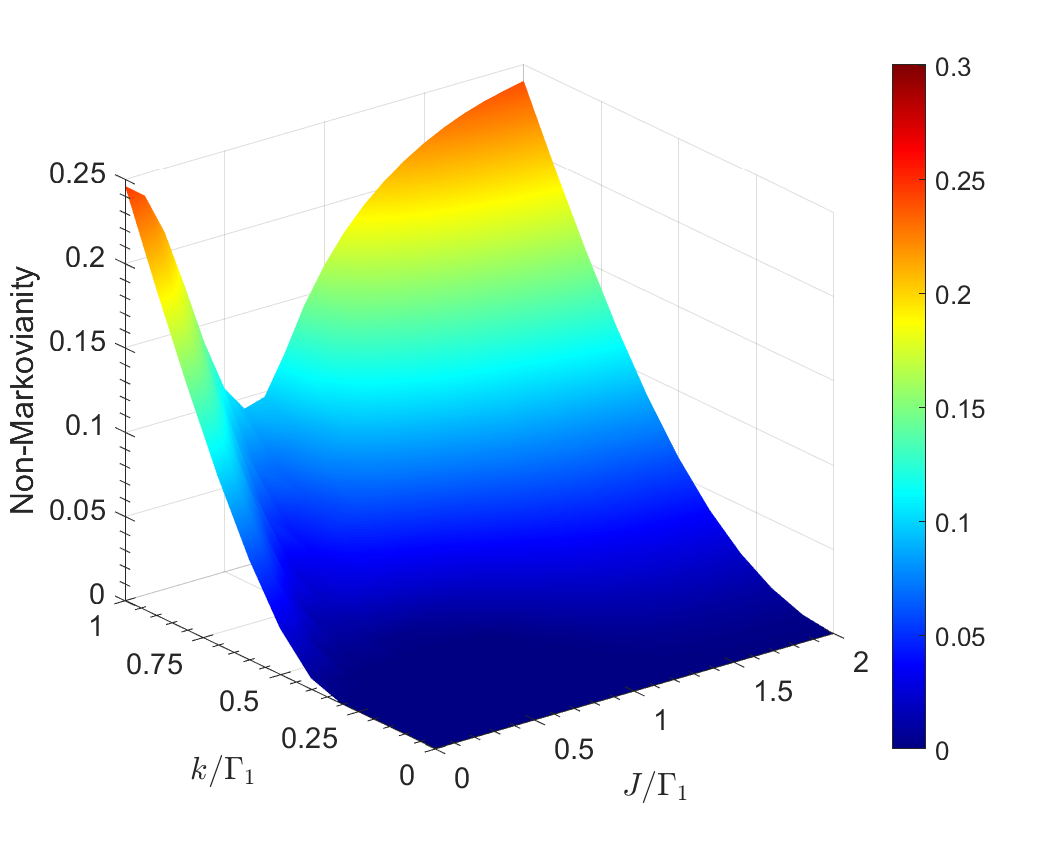}
    \centering
    \caption{The degree of non-Markovianity as functions of atom-pseudomode coupling $k/\Gamma_1$ and pseudomode-pseudomode coupling $J/\Gamma_1$ with $\Gamma_2/\Gamma_1=0.5$ and $\delta/\Gamma_1=0$.
}\label{Fig2}
\end{figure}

In Fig. \ref{Fig2}, the non-Markovianity \eqref{non} has been plotted as functions of atom-pseudomode coupling $k/\Gamma_1$ and pseudomode-pseudomode coupling $J/\Gamma_1$. From this figure, it can be seen that in the absence of pseudomode coupling strength ($J=0$), the evolution is Markovian in a weak coupling regime $k/\Gamma_1<0.25$ and non-Markovian in a strong coupling regime $k/\Gamma_1>0.25$.  In general, the non-Markovian feature of evolution is revealed for strong atom-pseudomode coupling for any pseudomode-pseudomode coupling strength.
Therefore, Fig. \ref{Fig2} clearly shows the regions in which the Markovian and non-Markovian dynamics take place.

Interestingly, by employing Eqs. \eqref{QSL} and \eqref{non}, it is possible to connect QSL time and degree of non-Markovianity as
\begin{equation}\label{QSL1}
\tau_{\mathrm{QSL}}=\tau \times \left(\frac{2\mathcal{N}}{1-|\widetilde{h}(\tau)|^2}+1\right)^{-1}.
\end{equation}

From Eq. \eqref{QSL1},  if the evolution is Markovian ($\mathcal{N}=0$), then the actual evolution time is equal to the QSL time, so we have a saturated bound for the QSL time. On the other hand, in a situation where the evolution is non-Markovian ($\mathcal{N}\neq 0$), the QSL time is shorter than the actual evolution time. 

\begin{figure}[t]
    \centering
  \includegraphics[width = 0.85\linewidth]{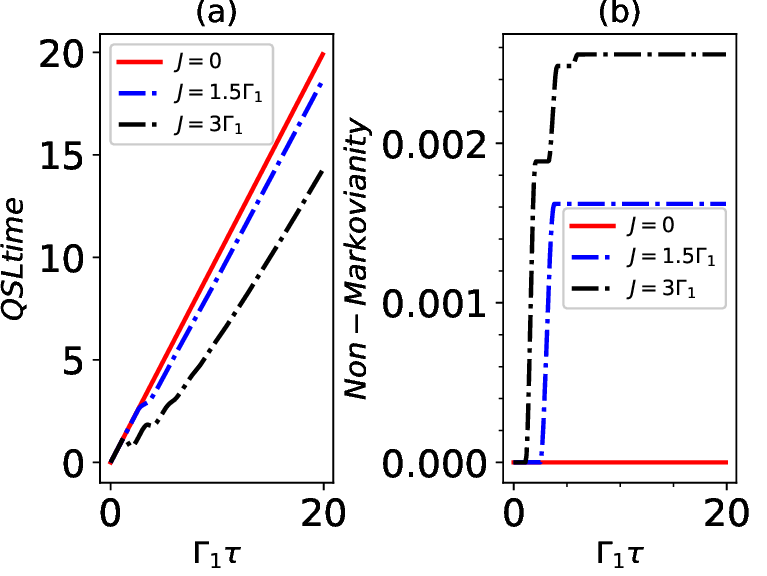}
    \centering
    \caption{(a) QSL time and (b) non-Markovianity as a function of $\Gamma_1 \tau$ in a weak coupling regime ($k/\Gamma_1=0.1$) for different values of $J/\Gamma_1$ with $\Gamma_2/\Gamma_1=0.5$ and $\delta/\Gamma_1=0$.
}\label{Fig3}
\end{figure}

In Fig. \ref{Fig3}, the QSL time and non-Markovianity have been plotted as a function of actual evolution time $\tau$ in a weak coupling regime with $k/\Gamma_1=0.1$ for different values of pseudomode-pseudomode coupling $J/\Gamma_1$. From Fig. \ref{Fig3}(a), one can see that the QSL time decreases with increasing $J/\Gamma_1$. Besides, Fig. \ref{Fig3}(b) shows that in this weak coupling regime, the non-Markovianity feature of the evolution is revealed by increasing $J/\Gamma_1$. Thus, we conclude that the QSL time decreases with the emergence of the non-Markovian feature of quantum evolution. Notably, one can also see that in Markovian evolution (red solid line), the QSL time is equal to the actual evolution time ($\tau_{\mathrm{QSL}}=\tau$) because $\mathcal{N}=0$, whose root is in Eq. \eqref{QSL1}. In other words, it can be said that the reduction of QSL time originates from the non-Markovian feature of the quantum evolution.
\begin{figure}[t]
    \centering
  \includegraphics[width = 0.85\linewidth]{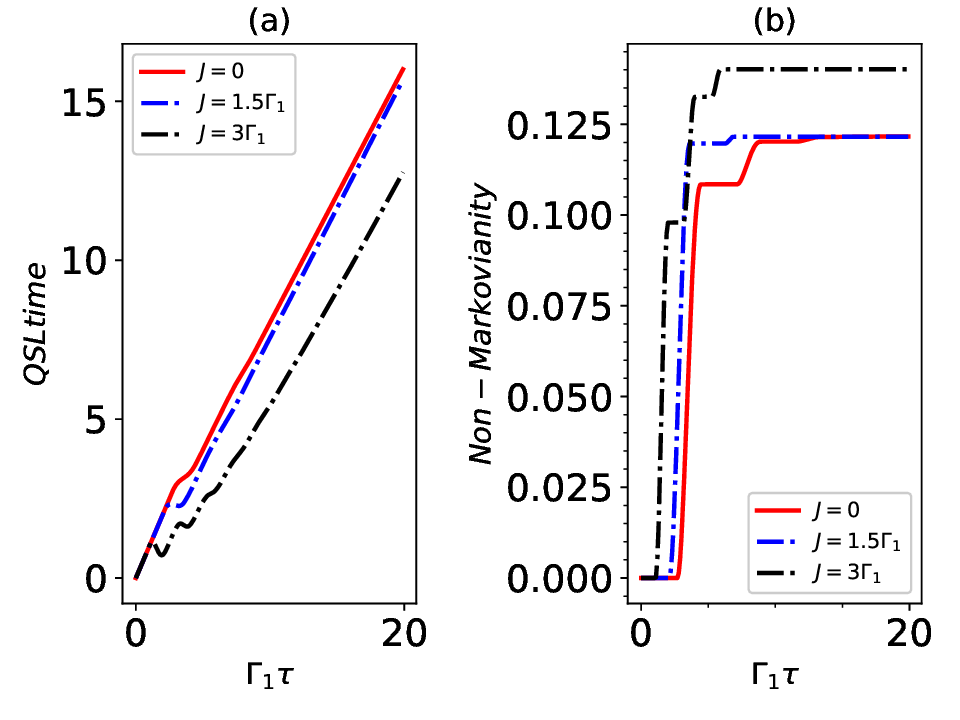}
    \centering
    \caption{(a) QSL time and (b) non-Markovianity as a function of $\Gamma_1 \tau$ in a strong coupling regime ($k/\Gamma_1=0.75$) for different values of $J/\Gamma_1$ with $\Gamma_2/\Gamma_1=0.5$ and $\delta/\Gamma_1=0$.
}\label{Fig4}
\end{figure}

Fig. \ref{Fig4} shows the QSL time and non-Markovianity as a function of actual evolution time $\tau$ in a strong coupling regime with $k/\Gamma_1=0.75$ for different values of $J/\Gamma_1$. It seems that the results are similar to what we observed in the weak coupling regime, however,  the non-Markovian feature of evolution can be observed in this case even when $J/\Gamma_1$ is zero, which is in good agreement with the results of Fig. \ref{Fig2}.
\begin{figure}[t]
    \centering
  \includegraphics[width = 0.85\linewidth]{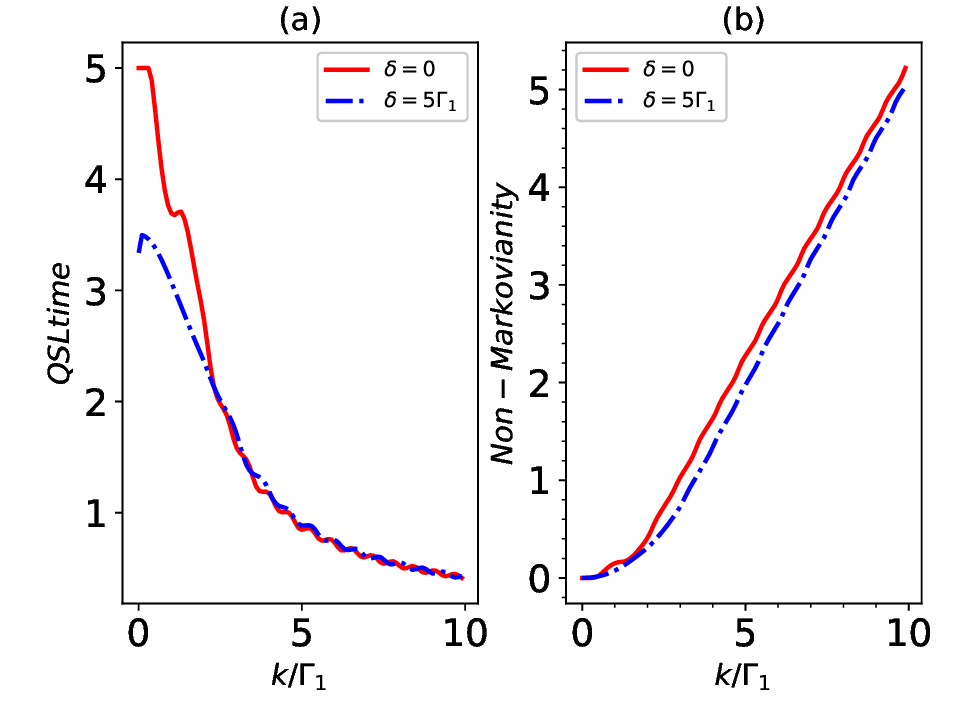}
    \centering
    \caption{(a) QSL time and (b) non-Markovianity as a function of $k/\Gamma_1$ for zero and nonzero values of $\delta/\Gamma_1$ with $\Gamma_2/\Gamma_1=0.5$, $J/\Gamma_1=1$, and $\tau=5$.
}\label{Fig5}
\end{figure}

In Fig. \ref{Fig5}(a), the QSL time has been plotted as a function of  $k/\Gamma_1$ for both resonant ($\delta/\Gamma_1=0$) and off-resonant ($\delta/\Gamma_1=5$) cases. The QSL time decreases with increasing $k/\Gamma_1$ for both resonant and off-resonant cases. Notice that the QSL time in the off-resonant case is shorter than in the resonant case. In Fig. \ref{Fig5}(b), we plot the non-Markovianity versus $k/\Gamma_1$. From this plot, we observe that the degree of non-Markovianity increases by the strengthening of the coupling between the atom and the first pseudomode for both resonant and off-resonant cases. By comparing Figs. \ref{Fig5}(a) and  \ref{Fig5}(b), it can be concluded that there exists an inverse relation between QSL time and non-Markovianity.    
\begin{figure}[t]
    \centering
  \includegraphics[width = 0.85\linewidth]{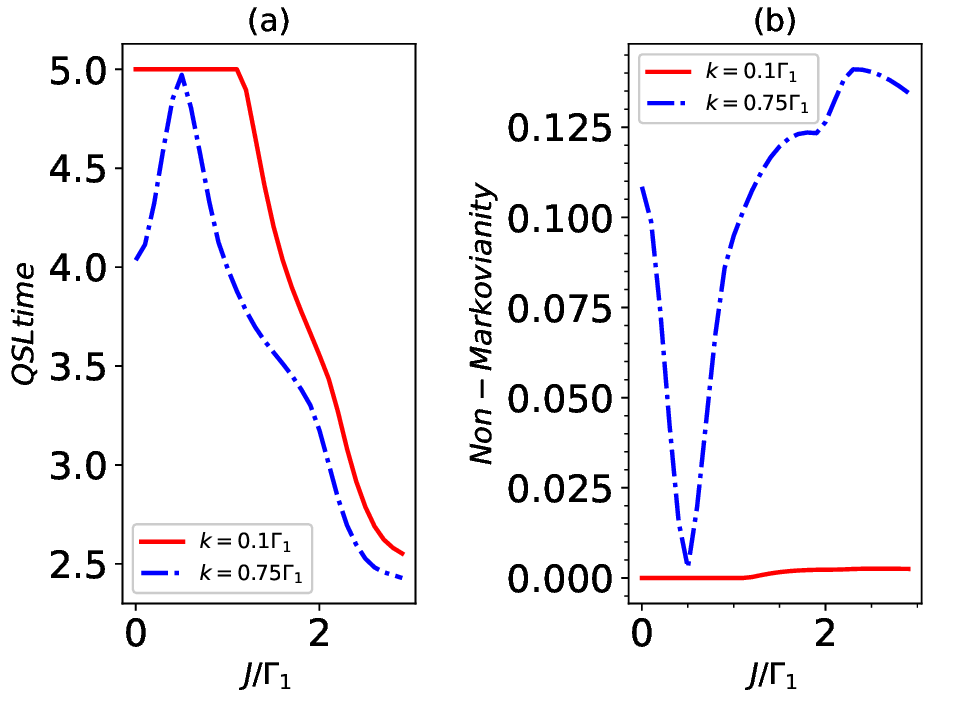}
    \centering
    \caption{(a) QSL time and (b) non-Markovianity as a function of $J/\Gamma_1$ for two different values of $k/\Gamma_1$ with $\Gamma_2/\Gamma_1=0.5$, $\delta/\Gamma_1=0$, and $\tau=5$.
}\label{Fig6}
\end{figure}

The QSL time and non-Markovianity have been plotted as a function of $J/\Gamma_1$ for two different values of $k/\Gamma_1$ in Fig. \ref{Fig6}. We see that in a strong coupling regime ($k/\Gamma_1=0.75$), the QSL time is shorter than that of the weak coupling regime ($k/\Gamma_1=0.1$). On the other hand, the non-Markovianity gets its maximum value in a strong coupling regime. Again, we find that there is an inverse relation between non-Markovianity and QSL time. These results are in good agreement with the previous figures. 
\begin{figure}[t]
    \centering
  \includegraphics[width = 0.85\linewidth]{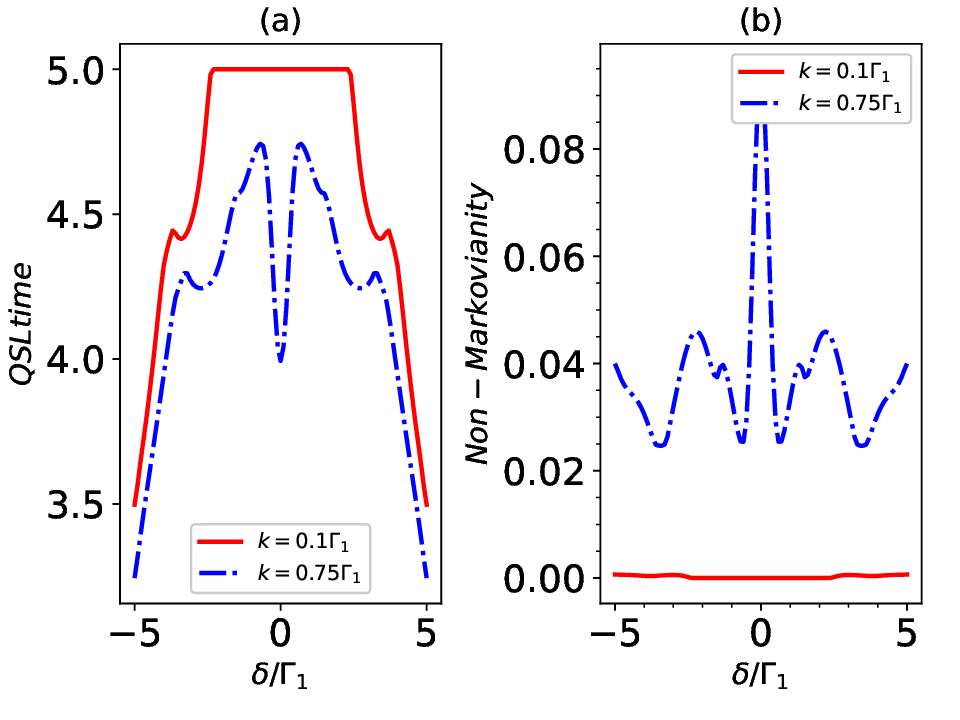}
    \centering
    \caption{(a) QSL time and (b) non-Markovianity as a function of $\delta/\Gamma_1$ for two different values of $k/\Gamma_1$ with $\Gamma_2/\Gamma_1=0.5$, $J/\Gamma_1=1$, and $\tau=5$.
}\label{Fig7}
\end{figure}

Finally, the QSL time and non-Markovianity have been illustrated as a function of the detuning parameter $\delta/\Gamma_1$ for two different values of $k/\Gamma_1$ in Fig. \ref{Fig7}. It is worth noting that the results are similar to the findings shown in the previous plots. Remarkably, the existence of an inverse relation between non-Markovianity and QSL time is also confirmed in this case. 

Note that the ratio between the QSL time, $\tau_{\mathrm{QSL}}$, and the actual time, $\tau$, quantifies the possibility of speeding up the quantum evolution. QSL bounds are saturated with no potential for speedup when $\tau_{\mathrm{QSL}}/\tau=1$. However, the QSL bound is unsaturated if $\tau_{\mathrm{QSL}}/\tau<1$, and there is potential for speeding up the evolution of the dynamical system. Our results show that $\tau_{\mathrm{QSL}}/\tau\leq1$, indicating that both of these scenarios occurred in our study.

\section{Conclusion and outlook}\label{conclusion}
Using a cavity-based engineered environment, we explored the possibility of manipulating the QSL time. In particular, we have observed that by constructing the Markovian environments consisting of two cavities, one can switch between quantum speed-down and speedup regimes for the dynamics of an artificial atom interacting with a pseudomode. Based on memory effects, the dynamics of the open systems are divided into two categories: memory-less (Markovian) and with-memory (non-Markovian) dynamics. Notably, we observed that there exists an inverse relation between the QSL time and non-Markovianity. In other words, the non-Markovianity feature of quantum evolution can lead to quantum speedup in the dynamics. Therefore, this feature leads to an intrinsic interest in the context of controlling the memory effects that are associated with open systems. The parameters included atom-pseudomode coupling, pseudomode-pseudomode coupling, and detuning between atom and pseudomode were used to manipulate the non-Markovianity. We found that in a strong coupling regime, for both atom-pseudomode and pseudomode-pseudomode couplings, the evolution is non-Markovian and the quantum speedup is revealed. As a result of this study, new insights are gained into the control of QSL time for open systems, which opens the door to further experimental development towards the development of quantum thermodynamic devices and quantum computers. 

\section*{CRediT authorship contribution statement}
\textbf{Maryam Hadipour:}  Writing--original draft,  Investigation. \textbf{Soroush Haseli:}  Writing--original draft, Writing--review \& editing, Supervision, Software, Investigation, Conceptualization.
\textbf{Saeed Haddadi:} Writing--review \& editing, Investigation, Formal analysis.

\section*{Acknowledgements}
 S. Haddadi was supported by Semnan University under Contract No. 21270.
 
\section*{Disclosures}
The authors declare that they have no known competing financial interests.

\section*{Data availability}
No datasets were generated or analyzed during the current study.



\begin{thebibliography}{99}
\bibitem{Deffner2017}Deffner S and Campbell S 2017 J. Phys. A: Math. Theor. 50, 453001.

\bibitem{Hilgevoord2002}Hilgevoord J 2002 Am. J. Phys. 70, 301.

\bibitem{Deffner2017p}Deffner S 2017 New J. Phys. 19, 103018.

\bibitem{Pfeifer1993}Pfeifer P 1993 Phys. Rev. Lett. 70, 3365.

\bibitem{Bukov2019}Bukov M, Sels D and Polkovnikov A 2019 Phys. Rev. X 9, 011034.

\bibitem{Funo2019}Funo K, Shiraishi N and Saito K 2019 New J. Phys. 21, 013006.

\bibitem{Hegerfeldt2013}Hegerfeldt G C 2013 Phys. Rev. Lett. 111, 260501.

\bibitem{Garcia-Pintos2019}Garcia-Pintos L P and del Campo A 2019 New J. Phys. 21, 033012.


\bibitem{Garcia-Pintos2019p}Garcia-Pintos L P and del Campo A 2019 New J. Phys. 21, 033012.

\bibitem{Kobe1994}Kobe D H and Aguilera-Navarro V C 1994 Phys. Rev. A 50, 933.

\bibitem{Jones2010}Jones P J and Kok P 2010 Phys. Rev. A 82, 022107.

\bibitem{Xu2016}Xu Z-Y 2016 New J. Phys. 18, 073005.

\bibitem{Deffner2013a}Deffner S and Lutz E 2013 J. Phys. A: Math. Theor. 46, 335302.

\bibitem{Shao2020}Shao Y, Liu B, Zhang M, Yuan H and Liu J 2020 Phys. Rev. Res. 2, 023299.

\bibitem{Russell2014}Russell B and Stepney S 2014 Phys. Rev. A 90, 012303.

\bibitem{Hu2020}Hu X, Sun S and Zheng Y 2020 Phys. Rev. A 101, 042107.

\bibitem{Cheneau2012}Cheneau M et al 2012 Nature 481, 484.

\bibitem{Campo2021}del Campo A 2021 Phys. Rev. Lett. 126, 180603.

\bibitem{Taddei2013}Taddei M M, Escher B M, Davidovich L and de Matos Filho R L 2013 Phys. Rev. Lett. 110, 050402.

\bibitem{Escher2011}Escher B M, de Matos Filho R L and Davidovich L 2011 Nat. Phys. 7, 406.

\bibitem{Lam2021}Lam M R et al 2021 Phys. Rev. X 11, 011035.

\bibitem{Poggi2019}Poggi P M 2019 Phys. Rev. A 99, 042116.

\bibitem{Caneva2009}Caneva T, Murphy M, Calarco T, Fazio R, Montangero S, Giovannetti V and Santoro G E 2009 Phys. Rev. Lett. 103, 240501.

\bibitem{Zhang2021}Zhang Y-J, Wei H, Yan W-B, Man Z-X, Xia Y-J and Fan H 2021 New J. Phys. 23, 113004.

\bibitem{Marvian2015}Marvian I and Lidar D A 2015 Phys. Rev. Lett. 115, 210402.

\bibitem{Bai2020}Bai S Y and An J H 2020 Phys. Rev. A 102, 060201.

\bibitem{Campaioli2017}Campaioli F, Pollock F A, Binder F C, Celeri L, Goold J, Vinjanampathy S and Modi K 2017 Phys. Rev. Lett. 118, 150601.

\bibitem{Hovhannisyan2013}Hovhannisyan K V, Perarnau-Llobet M, Huber M and Acın A 2013 Phys. Rev. Lett. 111, 240401.

\bibitem{Binder2015}Binder F C, Vinjanampathy S, Modi K and Goold J 2015 New J. Phys. 17, 075015.

\bibitem{OkuyamaMand2018}Okuyama M and Ohzeki M 2018 Phys. Rev. Lett. 120, 070402.

\bibitem{Shanahan2018}Shanahan B, Chenu A, Margolus N and del Campo A 2018 Phys. Rev. Lett. 120, 070401.

\bibitem{Bolonek-Lason2021}Bolonek-Lason K, Gonera J and Kosinski P 2021 Quantum 5, 482.

\bibitem{Poggi2021}Poggi P M, Campbell S and Deffner S 2021 PRX Quantum 2, 040349.

\bibitem{arXiv:2004.03078}Campaioli F, Yu C-s, Pollock F A and Modi K 2022  New J. Phys. 24, 065001.

\bibitem{Mandelstam1991}Mandelstam L and Tamm I 1991 \textit{The uncertainty relation between energy and time in non-relativistic quantum mechanics}
(Selected Papers, Berlin: Springer pp. 115).

\bibitem{Margolus1998}Margolus N and Levitin L B 1998 Physica D 120, 188.

\bibitem{Levitin2009}Levitin L B and Toffoli T 2009 Phys. Rev. Lett. 103, 160502.

\bibitem{Jozsa1994}Jozsa R 1994 J. Mod. Opt. 41, 2315.

\bibitem{Uhlmann1976}Uhlmann A 1976 Rep. Math. Phys. 9, 273.

\bibitem{Luo2004}Luo S and Zhang Q 2004 Phys. Rev. A 69, 032106.

\bibitem{Wu2020}Wu S-x and Yu C-s 2020 Sci. Rep. 10, 5500.

\bibitem{Deffner2013}Deffner S and Lutz E 2013 Phys. Rev. Lett. 111, 010402.

\bibitem{Wu2018}Wu S-x and Yu C-s 2018 Phys. Rev. A 98, 042132.

\bibitem{Cai2017}Cai X and Zheng Y 2017 Phys. Rev. A 95, 052104.

\bibitem{del2013}del Campo A, Egusquiza I L, Plenio M B and Huelga S F 2013 Phys. Rev. Lett. 110, 050403.

\bibitem{Ektesabi2017}Ektesabi A, Behzadi N and Faizi E 2017 Phys. Rev. A 95, 022115.

\bibitem{Liu2016}Liu H B, Yang W L, An J H and Xu Z Y 2016 Phys. Rev. A 93, 020105.

\bibitem{Breuer2007}H.-P. Breuer and F. Petruccione, \textit{The Theory of Open Quantum Systems} (Oxford University Press, Oxford, UK: Cambridge University Press, 2007).

\bibitem{Davies1976}Davies, E.B.: \textit{Quantum theory of open systems} (Academic Press, London 1976).

\bibitem{Wolf2008} Wolf, M M, Eisert J, Cubitt T S and Cirac J I 2008 Phys. Rev. Lett. 101, 150402. 

\bibitem{Breuer2009} Breuer H P, Laine E M and Piilo J 2009 Phys. Rev. Lett. 103, 210401. 

\bibitem{Rivas2010} Rivas A, Huelga S F and Plenio M B 2010 Phys. Rev. Lett. 105, 050403. 

\bibitem{Luo2012} Luo S, Fu S and Song H 2012 Phys. Rev. A 86, 044101. 

\bibitem{Zeng2011} Zeng H S, Tang N, Zheng Y P and Wang G Y 2011 Phys. Rev. A 84, 032118. 

\bibitem{He2017} He Z, Zeng H S, Li Y, Wang Q and Yao C 2017 Phys. Rev. A 96, 022106.

\bibitem{Zhang2014}Zhang Y J, Han  W, Xia Y J, Cao J P and Fan H 2014 Sci. Rep. 4, 4890. 

\bibitem{Cimmarusti2015}Cimmarusti A D, Yan Z, Patterson B D, Corcos L P, Orozco L A and Deffner S 2015 Phys. Rev. Lett. 114, 233602.

\bibitem{Zhang2015}Zhang Y J, Han W, Xia Y J, Cao J P and Fan H 2015 Phys. Rev. A 91, 032112.

\bibitem{Cianciaruso2017}Cianciaruso M, Maniscalco S and Adesso G 2017 Phys. Rev. A 96, 012105.

\bibitem{Sun2015}Sun Z, Liu J, Ma J and Wang X 2015 Sci. Rep. 5, 8444.

\bibitem{Xu2019}Xu K, Zhang G F and Liu W M 2019 Phys. Rev. A 100, 052305.

\bibitem{Xu2014}Xu Z Y, Luo S, YangWL, Liu C and Zhu S 2014 Phys. Rev. A 89, 012307.

\bibitem{Teittinen2019}Teittinen J, Lyyra H and Maniscalco S 2019 New J. Phys. 21, 123041.

\bibitem{Berrada2018}  Berrada K 2018 Physica E 95, 6.

\bibitem{Shahri2023}Shahri Y,  Hadipour M, Haddadi S, Dolatkhah H, and Haseli S 2023 Phys. Lett. A 470, 128783.

\bibitem{Gholizadeh2023} Gholizadeh A,  Hadipour M,  Haseli S,  Haddadi S,  Dolatkhah H 2023 Commun. Theor. Phys. 75, 075101.


\bibitem{Hadipour2023} Hadipour M,  Haseli S,  Dolatkhah H,  Haddadi S,  Czerwinski A 2022 Photonics 9, 875.


\bibitem{Mazzola2009} Mazzola L,  Maniscalco S,  Piilo J,  Suominen K A, and  Garraway B M 2009 Phys. Rev. A 80, 012104.

\bibitem{Xu2021} Xu K, Zhu H J, Zhang G F, Liu W M 2021 Phys. Rev. E 104, 064143.

\bibitem{pseudomode1} Garraway B M 1997 Phys. Rev. A 55, 2290.

\bibitem{pseudomode2} Garraway B M 1997 Phys. Rev. A 55, 4636.

\bibitem{pseudomode3} Mazzola L, Maniscalco S, Piilo J, Suominen K A, and Garraway B M 2009 Phys. Rev. A 80, 012104.

\bibitem{pseudomode4} Pleasance G, Garraway B M, and Petruccione F 2020  Phys. Rev. Research 2, 043058.








\end{thebibliography}
\end{document}